\documentclass[jgrga]{AGUTeX}

\def\referee#1{{#1}}

\usepackage[modulo]{lineno}

\usepackage[dvips]{graphicx}
\authorrunninghead{SCHRIJVER ET AL.}

\titlerunninghead{Frequency of extreme solar events}

\authoraddr{C.J. Schrijver
Lockheed Martin Advanced Technology Center, 
3251 Hanover Street, Palo Alto CA94304, USA
(schrijver@lmsal.com)}

\authoraddr{J. Beer, EAWAG, Swiss Federal Institute of Aquatic Science and Technology, Postfach 611 CH-8600 Duebendorf, SWITZERLAND (beer@eawag.ch)}

\authoraddr{U. Baltensperger, 
Paul Scherrer Institute, Laboratory of Atmospheric Chemistry, CH-5232
Villigen PSI, SWITZERLAND (urs.baltensperger@psi.ch)}

\authoraddr{E.W. Cliver, Space Vehicles Directorate, Air Force Research Laboratory, Sunspot, NM 88349, USA (ecliver@nso.edu)}

\authoraddr{H.S. Hudson, Space Sciences Laboratory, University of California, Berkeley, CA 94720, USA (hhudson@ssl.berkeley.edu)}

\authoraddr{M. G{\"u}del, University of Vienna, Department of Astronomy, T{\"u}rkenschanzstr. 17, A-1180 Vienna, AUSTRIA (manuel.guedel@univie.ac.at)}

\authoraddr{K.G. McCracken, Institute for Physical Science and Technology, University of Maryland, MD, USA (jellore@hinet.net.au)}

\authoraddr{R.A. Osten, Space Telescope Science Institute, 3700 San Martin Drive, Baltimore, MD 21218, USA (osten@stsci.edu)}

\authoraddr{Th. Peter, ETH Z{\"u}rich, Institut f{\"u}r Atmosph{\"a}re und Klima, CHN O 12.1, Universit{\"a}tstrasse 16, 8092 Z{\"u}rich, SWITZERLAND (thomas.peter@env.ethz.ch)}

\authoraddr{D.R. Soderblom, Space Telescope Science Institute, 3700 San Martin Drive, Baltimore, MD 21218, USA (drs@stsci.edu)}

\authoraddr{I.G. Usoskin, Sodankyl{\"a} Geophysical Observatory (Oulu unit), P.O. Box 3000, FIN-90014, University of Oulu, FINLAND (ilya.usoskin@oulu.fi)}

\authoraddr{E.W. Wolff, British Antarctic Survey, High Cross Madingley Road, Cambridge, CB3 0ET, UNITED KINGDOM (ewwo@bas.ac.uk)}

\def\ga{\mathrel{\mathchoice {\vcenter{\offinterlineskip\halign{\hfil
$\displaystyle##$\hfil\cr>\cr\sim\cr}}}
{\vcenter{\offinterlineskip\halign{\hfil$\textstyle##$\hfil\cr>\cr\sim\cr}}}
{\vcenter{\offinterlineskip\halign{\hfil$\scriptstyle##$\hfil\cr>\cr\sim\cr}}}
{\vcenter{\offinterlineskip\halign{\hfil$\scriptscriptstyle##$\hfil\cr>\cr\sim\cr}}}}}

\begin{document}


%
%

\title{Estimating the frequency of extremely energetic solar events,
  based on solar, stellar, lunar, and terrestrial records}
%

%
%



\authors{C. J. Schrijver\altaffilmark{1},
J. Beer\altaffilmark{2},
U. Baltensperger\altaffilmark{3},
E.W. Cliver\altaffilmark{4},
M. G{\"u}del\altaffilmark{5}, 
H.S. Hudson\altaffilmark{6}, 
K.G. McCracken\altaffilmark{7}, 
R.A. Osten\altaffilmark{8},
Th. Peter\altaffilmark{9}, 
D.R. Soderblom\altaffilmark{8}, 
I.G. Usoskin\altaffilmark{10}, and 
E.W. Wolff\altaffilmark{11}}

\altaffiltext{1}{Lockheed Martin Advanced Technology Center, 
Palo Alto, CA,  USA}
\altaffiltext{2}{Swiss Federal Inst. of Aquatic Science and Techn, D{\"u}bendorf, Switzerland}
\altaffiltext{3}{Paul Scherrer Institute,  Villigen, Switzerland}
\altaffiltext{4}{Space Vehicles Directorate, AFRL, Sunspot, NM, USA}
\altaffiltext{5}{University of Vienna, Austria}
\altaffiltext{6}{Space Science Laboratory, UC Berkeley, CA, USA}
\altaffiltext{7}{IPST, U. of Maryland, College Park, MD, USA}
\altaffiltext{8}{Space Telescope Science Institute, Baltimore, MD, USA}
\altaffiltext{9}{ETH Zuerich, Switzerland}
\altaffiltext{10}{Sodankyl\"a Geophysical Observatory and Department of Physics, University of Oulu, Finland}
\altaffiltext{11}{British Antarctic Survey, Cambridge, United Kingdom}

%
%


\begin{abstract}
  The most powerful explosions on the Sun --~in the form of bright
  flares, intense storms of solar energetic particles (SEPs), and fast
  coronal mass ejections (CMEs)~-- drive the most severe space-weather
  storms. Proxy records of flare energies based on SEPs in principle
  may offer the longest time base to study infrequent large events. We
  conclude that one suggested proxy, nitrate concentrations in polar
  ice cores, does not map reliably to SEP events. Concentrations of
  select radionuclides measured in natural archives may prove useful
  in extending the time interval of direct observations up to ten
  millennia, but as their calibration to solar flare fluences depends
  on multiple poorly known properties and processes, these proxies
  cannot presently be used to help determine the flare energy
  frequency distribution. Being thus limited to the use of direct
  flare observations, we evaluate the probabilities of large-energy
  solar explosions by combining solar flare observations with an
  ensemble of stellar flare observations. \referee{We conclude that
    solar flare energies form a relatively smooth distribution from
    small events to large flares, while flares on magnetically-active,
    young Sun-like stars have energies and frequencies markedly in
    excess of strong solar flares, even after an empirical scaling
    with the mean activity level of these stars. In order to
    empirically quantify the frequency of uncommonly large solar
    flares extensive surveys of stars of near-solar age need to be
    obtained, such as is feasible with the Kepler satellite. Because
    the likelihood of flares larger than approximately X30 remains
    empirically unconstrained, we present indirect arguments, based on
    records of sunspots and on statistical arguments, that solar
    flares in the past four centuries have likely not substantially
    exceeded the level of the largest flares observed in the space
    era, and that there is at most about a 10\%\ chance of a flare
    larger than about X30 in the next 30 years.}
\end{abstract}

%
%

%

\begin{article}

%
%

\section{Introduction}
The Sun displays explosive and eruptive phenomena that span a range of
at least a factor of $10^8$ in energy, from the present-day detection
limits for ``nanoflares'' and the eruptions of small fibrils up to
large, highly-energetic ``X-class'' flares and coronal mass
ejections. At the lowest energies, millions of such events occur each
day above the detection limit of $\sim 10^{24}$\,ergs. The largest
observed solar flares, with energies substantially exceeding
$10^{33}$\,ergs, occur as infrequently as once per decade or less.

\begin{figure}
\noindent\includegraphics[width=8.4cm,trim=30mm 160mm 100mm -40mm,clip]{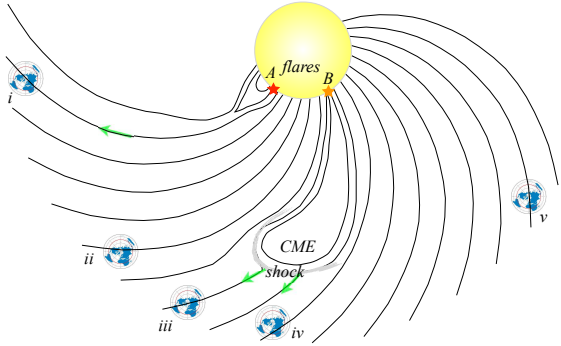}
\label{fig:1}
\caption{\em Illustration of visibility and propagation of solar
  explosive and eruptive events (modified after Reames, 1999; for an
  MHD simulation see, e.g., Rouillard et al., 2011), viewed from
  different orbital phases of Earth ({\em i - v\/}; not to scale)
  clarifying that flares (seen as photons) and energetic particle
  events are related, but that their detection depends on the
  evolution of the event on the Sun, on the heliospheric counterpart
  of any eruption (CME), and on the perspective from Earth (or a
  distant satellite, like the STEREO spacecraft). An eruption at
  position $A$ may be confined to the solar magnetic field, never
  reaching the heliosphere; depending on the magnetic geometry, solar
  energetic particles (SEP; green arrows) may escape into the
  heliosphere, following the Parker spiral of the field and detectable
  (as a rapid SEP event) near Earth only for orbital phases around
  {\em i\/}. An eruption at position $B$ could lead to a
  near-instantaneous SEP event for orbital phases around {\em iii\/}
  and later shock-accelerated (gradual) SEPs at phases {\em
    iii-iv\/}. Streaming-limit saturation of the SEP flux density
  would be observed for orbital phases {\em iii - iv\/}, with the SEP
  flux density exceeding the streaming limit when the shock front
  crosses Earth for phase {\em iv\/}.}
\end{figure}
Solar events have an increasing potential to impact mankind's
technological infrastructure with increasing flare energy, most
effectively in the range of X-class flares, i.e. from a few times
$10^{31}$\,ergs upward \cite[see,
e.g.,][]{severeswx2008,fema2010,kappenman2010,swximpactlloyds2011,jason2011}.

Solar flares are the observed brightenings that result from a rapid
conversion of energy contained in the electrical currents and in the
magnetic field within the solar corona into photons through a chain of
processes that involves magnetic reconnection, particle acceleration,
plasma heating, and ionization, eventually leading to electromagnetic
radiation. Large solar flares (defined here as involving energies in
excess of some $10^{31}$\,ergs) can accelerate particles to high
energies and are generally associated
with coronal mass ejections in which matter and magnetic field are
ejected into the heliosphere at velocities of up to  $\approx
3,000$\,km/s. The ejections often drive shocks in which more accelerated
particles are generated within the low corona and in the
heliosphere. Due to these processes, solar flares are frequently
associated with solar energetic particle (SEP) events near Earth (see,
e.g., the reviews by \cite{benz2008,schrijver2009b}). We discuss the
relationships between these and other aspects of solar and space
weather in some more detail in Sec.~\ref{sec:intro}.

Large solar events drive episodes of severe space weather, including
strong geomagnetic storms, enhanced particle radiation,
pronounced ionospheric perturbations, and
powerful geomagnetically-induced Earth currents, all of which affect our
technological infrastructure from communications to electric power
\citep{severeswx2008}. It is therefore of substantial interest to
establish the probability distribution for the largest solar flares
and their associated energetic particle events and coronal mass
ejections. 

Direct measurements of the energies involved in solar events have been
within the realm of the possible only since the beginning of the space
age. Whereas the instrumental record spans almost eight decades, it
begins with H$\alpha$ monitoring, with observation of flare ionizing
radiation and energetic particles (initially by indirect means) as
well as radio emission added over time, eventually culminating in
global solar coverage only since 2011 with a patchwork of passbands
that range from $\gamma$-rays to radio that can, with difficulty, be
linked into a comprehensive view of the energies involved
\citep[e.g.,][]{emslie+etal2004,emslie+etal2005}. Hence, the
frequencies of solar coronal storms that may occur only once per
century, or even less frequently, remain to be established.

As we have only a limited understanding of the formation of
magnetically-active solar regions and of their explosive potential, we
have no theoretical framework that can be used to extrapolate the
observed energy distribution of solar flares to energies that lie
beyond the observed range.  Sun-like stars provide evidence that
larger magnetic explosions are possible, with observed energies that
exceed the largest observed solar flares by at least three orders of
magnitude. But, as we discuss in later sections, such stars are
typically much younger and thus magnetically much more active than the
present-day Sun, and with generally different patterns in their
dynamos as reflected, for example, in the existence of high-latitude
or polar activity and in the general lack of simple cycle signatures
\citep[e.g.,][]{2005LRSP....2....8B,hall_2008}. Can the Sun still
power events substantially larger than, say, a large, infrequent X30
flare, and, if it can, how likely are such events?  How likely are
solar energetic particle events of various magnitudes?

In this study, we evaluate and integrate the available evidence 
to quantify the frequency distribution of the most energetic solar
events. To this end, we combine direct observations of photons emitted
by solar flares with those of their stellar counterparts. Such a
comparison offers the advantage that observing an ensemble of
Sun-like stars enables us to collect statistics on the equivalent of
thousands of years of solar time, albeit subject to the problem that
stellar flares are typically observed on stars that are much more active
than the Sun has been at any time in recent millennia. 

\nocite{1999SSRv...90..413R,2011ApJ...735....7R} The association of
solar flaring and frequent attendant CMEs with energetic particle
events offers complementary sources of information on the statistics
of extreme solar coronal storms.  First, energetic particles leave
observable signatures when they cause nuclear reactions in rocks that
are exposed to them, such as lunar rocks \citep{2009GeCoA..73.2163N},
and even in terrestrial rocks that are protected by the Earth's
magnetosphere and atmosphere.  Second, such energetic particles induce
nuclear reactions in the terrestrial atmosphere which leave
radioactive fingerprints in a variety of forms, including cosmogenic
radionuclides $^{14}$C and $^{10}$Be, that can be traced in the
geosphere as deposited, e.g., in polar ice or in trees.  Third, the
particles impacting the Earth's upper atmospheric layers are expected
to cause shifts in the chemical balance which may leave identifiable
signatures in precipitation records; in particular, this pathway to
long-term records on extreme solar events has been suggested for
nitrate concentrations in polar ice (Sect.~\ref{sec:nitrates}).

Each of these indirect measures (which we discuss in
Sects.~\ref{sec:nuclides} and~\ref{sec:nitrates}) offers its own
difficulties related to its specific geochemical properties and the
transport from the atmosphere into its archive. For example, $^{14}$C
forms CO$_2$ and enters the global carbon cycle where it becomes
heavily smoothed in time; $^{10}$Be spikes are subject to fluctuations
of the climate and weather, both on Earth and throughout the
heliosphere; exposed rock faces can only tell us about the cumulative
effect of solar energetic particles over the lesser of the decay time
of the radionuclides involved and the duration over which a rock face
is exposed to solar particles. All of these radionuclide records sit
on top of a background that is associated with galactic cosmic rays,
which themselves are modulated on time scales upward of a few years by
the variable solar wind, the heliospheric magnetic field, and the
terrestrial magnetic field.  Chemical signatures, as we discuss below,
offer even greater difficulties, and we  conclude that we  do not
currently see a way to use nitrate concentrations as indicators of SEP events.

In addition to these challenges in understanding the temporal
modulations and integration of the records of solar energetic
particles, there are challenges related to the creation and
propagation of these particles before they are recorded. The relative
importance of flares and CME-driven shocks for large SEP events
continues to be debated: SEPs are generated both during the
initial phases of a flare and in the propagation of CME shocks into
and through the heliosphere. Line-of-sight photons and magnetically
guided SEPs follow distinct pathways to Earth, so that flares and SEP
events at Earth may be poorly correlated in time, contributing to a
complex statistical relationship between the phenomena. Establishing
their relationship requires that we understand the angular widths of the particle
distributions entering into, and generated within, the heliosphere
compared to the $2\pi$ solid angle available to flare photons. Another
complication, yet to be properly understood, involves the propagation
of the SEPs through the heliosphere, which appears subject to a
saturation effect referred to as the ``streaming limit''
(Sect.~\ref{sec:streaming}). Some of the geometrical considerations
involved in the flare-SEP correlation in observations at Earth are
illustrated in Fig.~\ref{fig:1}. These and other issues are discussed
in subsequent sections in the context of the available literature.

Transport of energetic particles in the geomagnetic field and
atmosphere, including a nuclear atmospheric cascade/shower, is
relatively well understood \citep[e.g.,][]{vainio09}. Whereas the
transport of galactic cosmic rays (i.e., energetic particles
originating outside the heliosphere) through the heliosphere is
relatively well understood
\citep[e.g.,][]{jokipii00,potgieter01,2004JGRA..10901101C}, the
propagation of solar energetic particles --~sometimes called solar
cosmic rays~-- (i.e., those originating from a flare site or from a
heliospheric shock associated with a solar eruption) is subject to
substantial uncertainties (see Section~\ref{sec:streaming}). The
parameters that set the spectral shape of the particle energy
distribution are mostly empirically determined, adding additional
difficulties when seeking to quantify the most energetic events that
have been rarely observed, in particular for possible very rare events
that have never been observed directly at all.

In Section~\ref{sec:intro} we present a brief overview of the
connection between solar flares and energetic particles before they
enter the detection systems in the form of spacecraft, ground-based
detectors, rocks, ice, or biosphere.  This section is mainly meant for
readers who are relatively unfamiliar with these processes and their
terminologies.  After this brief introduction of some of the issues to
be dealt with when using photons and tracers of energetic particles to
learn about solar energetic events, we proceed to integrate solar,
stellar, lunar, and terrestrial records in our attempt to establish
the probability distribution of the largest solar energetic events.

Sections~\ref{sec:nuclides} and~\ref{sec:nitrates} lead to the finding
that SEP records cannot be used to put tight constraints on the
statistics of the largest solar flares, at least at present.  The use
of cosmogenic radionuclides to constrain SEPs near Earth is discussed
in Section~\ref{sec:nuclides}. In Section~\ref{sec:nitrates} we review
the evidence, obtained in conjuction with this study, that nitrate
concentrations in ice deposits cannot, at present, be used to learn
about SEP events because the analyses of multiple ice cores has
recently cast doubt on the suggestion that spikes in nitrate
concentrations correlate with SEP events; ice nitrate concentrations
may yet be validated as a quantitative metric for SEP events, but at
present, the correspondence needs to be viewed at most as possible.
Sections~\ref{sec:nuclides} and~\ref{sec:nitrates} 
clarify why, in the end, we have to rely on direct
observations of flares. These two sections discuss constraints on the
flare energy frequency distribution that turn out to be weak at best;
they could be skipped on first reading.

Solar and stellar observations do provide interesting information on 
the flare energy distribution over many orders of
magnitude: the comparison of solar and stellar flare observations in
various segments of the electromagnetic spectrum is discussed in
Section~\ref{sec:flares}.

Section~\ref{sec:streaming} contains an evaluation of the
transformation of direct SEP and flare observations to a common scale
for the source strengths near the Sun.

Flares and eruptions take their energy from the magnetic field within
active regions; the implications of active-region sizes compared to
the energies involved in flares and CMEs are described in
Section~\ref{sec:conversion}.

We integrate the various findings in a discussion in
Section~\ref{sec:discussion}.

\section{Flares, CMEs, photons and energetic particles}\label{sec:intro}
``Flares'' are, by definition, relatively rapid brightenings in the
photon spectrum of the Sun and other stars. The signatures of flares
can be found from very high-energy $\gamma$-ray emission to km-wave
radio emissions. The bulk of a flare's energy is radiated at visible
wavelengths (see Section~\ref{sec:flares}), but because of the bright
background of the photospheric emission, flares have the highest
contrast at X-ray, EUV, and radio emissions. Consequently, solar flare
monitors generally report on the X-ray signature of the solar spectral
irradiance.

Flares on stars other than the Sun, involving, for example, fully
convective late-M type dwarf stars or somewhat evolved stars of near
solar mass in tidally locked binary systems, share many of the
characterizing properties of solar flares.  Stellar flares reported on
in the literature \citep[e.g.,][]{audard+etal2000a,2004A&ARv..12...71G,2007A&A...468..463S,walkowicz+etal2010} are generally much more energetic than even large
solar flares, but that is mostly because of the observational
constraints of having to measure these stellar flares against the
full-disk background coronal emission in stars that are X-ray bright,
i.e., typically young, rapidly-spinning stars compared to the rather
slowly rotating Sun \citep[e.g.,][]{guedel+etal2003}.

The thermal emission of flaring ranges 
from below a million degrees for the smallest events observed in
quiet-Sun ephemeral regions to at least 100\,MK for large-energy stellar events
(e.g., \citet{2007ApJ...654.1052O}; see Sect.~\ref{sec:stellarflares}
for a discussion of some of the largest stellar flares observed to date). In emissions
characteristic of high energies (providing direct or indirect
measurements of non-thermal particle populations or direct
measurements of high-temperature thermal emission), solar and stellar
flares alike show fast rise and exponential decay phases (sometimes
summarily characterized as ``FRED'').  As flares transition from the
impulsive (fast-rise) to the decay phase, the spectral irradiance
typically follows the so-called Neupert effect \citep{neupert1968,2002A&A...392..699V}: lower-energy emissions
(e.g., soft X-rays) behave, to first order, as the time integral of
high-energy emissions such as hard X-rays, non-thermal radio emission,
or near-UV (or U-band) emission (for some examples of the Neupert effect in stellar
flares, see \cite{2002ApJ...580L..73G} for the dM5.5 star Proxima
Centauri; \cite{1996ApJ...471.1002G} for the M5.5Ve star UV\,Ceti;
\cite{2003ApJ...597..535H} for the dMe star AD\,Leo; and
\cite{2004ApJS..153..317O} for the K1IV+G5IV binary HR\,1099).

Solar flares are typically characterized by the NOAA/GOES magnitude
scale which measures the peak brightness (increasing in orders of
magnitude as A, B, C, M, and X, each followed by a number from 0 to
9.9 measuring
the peak brightness within a decade). 
Many flares (often 'compact flares') are characterized
by impulsive brightenings and rapid decays, bringing most of the solar
spectral irradiance back to near-preflare levels within a matter of
minutes to 
tens of minutes; other ``long-duration flares'' can have a gradual rise
and decay, sometimes lasting more than a dozen hours. Not only are the
time scales different, the peak emissions occur from hard X-rays to
relatively long-wavelength EUV, shifting overall to lower energies during the decay phase of any
given flare, while differing between compact and eruptive flares,
and between
active-region flares and quiet-Sun filament eruptions that lead
to CMEs \citep[e.g.,][]{benz2008}. Consequently, the GOES classification scheme is not
unambiguously useful as a metric for total flare energies; we discuss
this problem in Section~\ref{sec:flares}. 

Whereas the distinct
appearance of flares of different magnitudes and phases of evolution
in different passbands complicates the bolometric calibration sought
in this study, it is likely to play a role in enabling us to detect 
stellar flares against the full-disk background. 
The fact that flares shift through X/(E)UV wavelengths as a
function of their magnitude and evolutionary phase restricts the
range of flare energies that shows up in any such passband;
this limits the ``depth'' of the distribution function, i.e., the
ratio of largest to smallest flare observable within a given passband 
\citep[e.g.][]{guedel+etal2003}, leaving the largest flares to stand out against the
relatively weakened composite
background. Even then, the ``background'' itself contains, and may be
dominated by, a composite of
flares, cf.\ the discussion by \cite{2003ApJ...589..983A} of a long
observation of the M-dwarf star binary UV\,Ceti
which shows continuous variability with no well-defined
non-flaring level. 

The broad wavelength range involved in solar and stellar flares makes
it hard to observe the bolometric behavior of flares directly, because 
observations are typically limited to a relatively narrow bandpass. Hence,
transforming the measured signal to an estimated bolometric fluence involves
rather uncertain transformations, as discussed in
Section~\ref{sec:flares}.

Whereas flare photons from Sun and stars are detectable
with present-day instrumentation, they leave no signatures that enable
us to look back in time. 
SEPs that impact Earth or other solar-system bodies do leave such signatures, but their
generation and transport introduce a range of challenges to be dealt
with before SEP signatures can be used to quantify the frequency spectrum of
solar flare energies.

Over 40 years ago, \cite{1970SoPh...12..266L} presented evidence that
there are two principal ways in which particles are accelerated at the
Sun: (1) a process associated with reconnection in solar flares that
has type III (fast-drift) radio bursts as its defining meter-wave
radio emission and electrons with energy of $\sim 10$\,keV as its
characteristic particle acceleration ; and (2) acceleration at a shock
wave manifested by a (slow-drift) type II metric burst, which is
thought to reflect acceleration of escaping electrons and protons at
all energies. \cite{1978SoPh...57..429K} suggested that the type~II
shocks associated with SEP events were driven by CMEs, a suggestion
that has found increasing support
\citep{2002ApJ...572L.103G,2004ApJ...605..902C,2005JGRA..11012S07G}.

By the mid-1980s the basic two-class picture of SEP acceleration was
strengthened by elemental-composition and charge-state measurements of
$^3$He and higher-mass ions.  The observations revealed that the
$^3$He and Fe abundances in the flare (type III) SEP events were
enhanced by about a factor of $10^3-10^4$ and 10, respectively,
relative to that in the shock (type II) events, and that Fe charge
states were characteristically higher in the flare events (around 20
in flares versus $\sim 11-14$ for the large SEP events associated with
shocks), see the review by \cite{1999SSRv...90..413R}.

The original “two-class” paradigm was challenged in the late 1990s
when several large (and therefore presumably shock-associated) SEP
events exhibited the elemental composition and charge states of the
flare/reconnection SEP events 
\citep[see][for a historical review of SEP research]{2009IAUS..257..401C}.  
Over time, these unusual large events
were interpreted \citep[e.g.,][]{2005ApJ...625..474T,2006ApJ...646.1319T} in terms of particle 
acceleration in quasi-perpendicular shocks of “remnant” seed particles
remaining in the low corona and inner heliosphere from earlier flares.  Around the
maximum of the solar cycle, when flares are most frequent, enhanced
$^3$He SEP populations are observed in situ near Earth some $60$\%\ of
the time \citep{2003AIPC..679..652W}.  It is presumed that these remnant populations are also
present near the Sun where they can be acted on by shocks.  Because
the remnant particles have the composition and charge state
characteristics of flare-accelerated particles, the resulting SEP
event looks like a high-energy flare-event, even though the ultimate
accelerator is a shock.

Ground-based neutron monitors and ionization chambers have observed
some 70 so-called ground-level enhancements (GLEs) in the past seven decades, 
indicating the presence of fluxes of ions in the energy range
$1<E<20$\,GeV, which will have produced radionuclides. If the
initiating solar activity was within about $45^\circ$ from central
meridian, however, the interplanetary CME will more strongly scatter
GCRs, resulting in temporary decrease in the GCR intensity at
Earth \citep{1942TeMAE..47..185L}, commonly referred to as a ``Forbush decrease''. 
The cosmogenic radionuclide formation at Earth may, in some cases, be  
overcompensated by the  Forbush decrease with an associated reduction
in GCRs by about 10\%\ for about a week \citep{usoskin08}, but the
details of that depend on the conditions of the event \citep{2004AdSpR..34..381R}. For example,
solar eruptions near the western limb produce the most intense GLEs,
and contain the highest fluxes of particles with energies in excess of
5\,GeV, while in this case there typically is no Forbush decrease at Earth.

We note that for SEP proxies with a long mixing time scale within the
Earth's atmosphere prior to deposition (specifically for the $^{10}$Be
concentration discussed below) there is the additional complicating
factor that SEP-induced increases in the proxy ride on top of
variations associated with the GCR variations that are associated with
variations in the heliospheric magnetic field and the 
solar wind on time scales of years or more. To
differentiate between, say, large-fluence SEP events and extended
cycle minima, one has to make assumptions about the heliosphere that
are difficult to validate.

\begin{figure}[t]
\noindent\includegraphics[width=8.4cm]{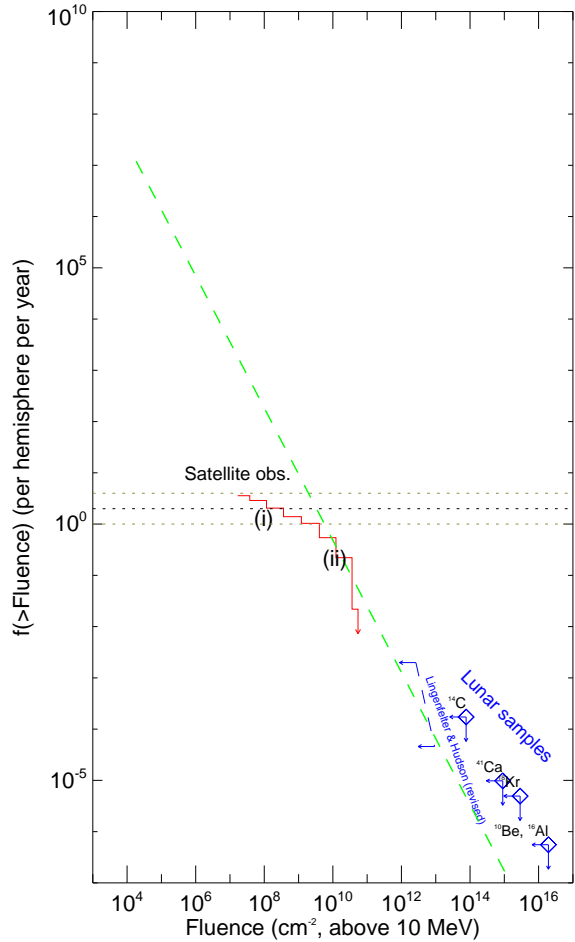}
\label{fig:2b}
\caption{\em Downward-cumulative frequency distribution
    (red) of fluences of solar energetic particle (SEP) events of
    fluences $F$ or larger (for particles with energies in excess of
    10\,MeV) for satellite observations (histogram), and for upper
    limits derived from lunar radionuclides (blue) and terrestrial
    records ($^{14}C$ upper limit shown by a (blue) dashed line). The
    green dashed line shows the slope of the fit to the flare energy
    frequency distribution from panel {\em a} for comparison, scaled
    to go through the kink in the satellite-based SEP fluence
    frequency distribution. Labels (i) and (ii) are discussed in the
    last paragraph of Sect.~\ref{sec:streaming}.}
\end{figure}
Within the heliosphere, SEP propagation may be subject to 
a ``streaming limit'' for particles
escaping from a shock acceleration region. This is a type of saturation effect
caused when protons streaming from the shock are hampered by their
propagation
in their own enhancement of the upstream
waves \citep{2010ApJ...723.1286R}, whose existence is an essential
component of the theory of diffusive shock acceleration.  This streaming limit 
does not apply near the shock, so SEP fluxes can exceed the
streaming limit when a shock passes directly over Earth, or over a
satellite outside the geomagnetic field.  

\section{Radionuclides as tracers of past solar energetic particle events}\label{sec:nuclides}
\subsection{Extraterrestrial radionuclides}\label{sec:extraterrestrial}
A direct way to determine the statistics of solar energetic particle
events is
to measure energetic particles  with space-based
instrumentation. A compilation of the fluences for such events for
particle energies exceeding 10\,MeV is shown as a red histogram in  Fig.~\ref{fig:2b}
(based on data from \citet{2001JGR...10621585M}). These data are
naturally limited to event frequencies exceeding once per fifty years,
as that is the current span of the observational record. On the
low-fluence side the range of accessible energies in the frequency 
spectrum is limited by the detection threshold
against the GCR background.

Some information on events that are much rarer than once per century
can be extracted from 'exposures' that have lasted much
longer than a few decades.  SEPs that impact solar-system bodies leave
traces in the form of a mixture of radioactive nuclides. 
The production of cosmogenic radionuclides from the exposure to SEPs
can be calculated for a specified elemental composition of the rock
and a given shape of the differential energy spectrum using Monte
Carlo simulations \citep{masarik94}. In a rock, only the time-integrated production rate
(a balance between production and decay) is recoverable. The integration
time depends on the half-life of the radionuclide in question. As a
consequence of the much steeper energy spectrum of SEPs compared to
that of GCRs, SEPs only produce cosmogenic radionuclides in the outermost
layers of the rocks. This differentiation between GCRs and SEPs as a
function of depth creates a natural spectrometer that enables
correction for the contribution from GCR-induced production, although
this does require assumptions on the SEP energy spectrum in
order to thereby estimate upper limits of the frequency of SEP events (see
\cite{usoskin_2008}, and references therein).

When rocky
material from the Moon is analyzed, we have access to the
cumulative dose of SEPs without the complicating factors of
terrestrial magnetic and atmospheric shielding or the effects of a
dynamic weathering environment. The combined results of lunar rock studies, compiled by
\cite{usoskin_2008}, assuming that the upper limits to the fluences are
associated with a few events over the isotopes' life time, are shown in
Fig.~\ref{fig:2b}. These upper limits emphasize 
the downturn seen at the high-fluence end of the frequency distribution of satellite
SEP observations, but they
are not particularly restrictive in establishing the shape of the
spectrum or the strength or fluence of a possible largest SEP event size.

\subsection{Cosmogenic radionuclides on Earth}\label{sec:terrestrial}
The combination of the SEP fluence frequencies measured by satellites
and the upper limits based on the analysis of lunar rocks shown in
Fig.~\ref{fig:2b} illustrates the need to fill a gap for
events with cumulative frequencies of less than once per few
decades. In this section, we discuss one possibility that is currently
being explored, which is the measurement of terrestrial radionuclides
stored in a stratified manner that enables setting tighter limits on
lower-fluence events.

Terrestrial cosmogenic radionuclides are produced mainly by
spallation-type nuclear
interactions between high-energy (GeV) particles and nuclei of the
dominant atmospheric constituents (N, O, Ar). After production, those
radionuclides that end up stored in naturally stratified ``archives''
such as ice deposits, trees, and sediments, prove most useful to our
purpose.

Records of cosmogenic radionuclides provide blended information about
the solar magnetic activity, the strength of the geomagnetic dipole
field, and atmospheric transport and deposition processes. By using
independent information about the geomagnetic dipole field, and by combining different
records of $^{10}$Be from ice cores with $^{14}$C from tree rings, a
rather clean signal of the variations in the GCRs due to varying
levels of solar activity can be extracted for at least the past
10,000 years. That record reveals the variability of the solar dynamo
and the associated heliospheric magnetic field on time scales ranging
from decades to millennia, with grand minima and maxima throughout the
long record \citep{solanki04,vonmoos06,usoskin07,steinhilber08,2010JGRA..11501104S,2011SSRv..tmp..324M}.

Not only the long-term variations can thus be recovered:  there is
some promise of recovering shorter-term spikes, albeit that these are washed out by
the transport process between generation and deposition, while set
against a variable background of the solar dynamo.  The time it takes to
transport a newly produced cosmogenic radionuclide from the atmosphere
into an archive depends mainly on the altitude at which it is
produced. This ranges from weeks for the troposphere to years for
the stratosphere \citep{raisbeck81,field06}. As a consequence, the production signal stored in
the archive is smoothed and the temporal resolution is limited to
about one year at best. The higher the desired temporal resolution,
the more the signal will be influenced by transport processes. Over the
past 5 years, the use of global circulation models (GCM) has greatly
improved our understanding of the manner in which atmospheric
transport processes influence the deposition of $^{10}$Be and other
radionuclides into polar ice \citep{field06,heikkila09}. 

To produce cosmogenic radionuclides a primary (galactic or solar) cosmic ray needs
energies above about 500\,MeV with a specific yield function depending
on the particular isotope. Because of the relatively low energies
in SEPs (compared to GCRs) the majority of them can only enter the
Earth's atmosphere at high magnetic latitudes (exceeding about
$60^\circ$). Moreover, again because of the relative softness of the
SEP energy spectrum, the contribution  to the cosmogenic isotope
production 
of most of the SEP events that
can be observed by satellites in orbit is too
small to be detected in ice, rocks, or biosphere against the background production of a
radionuclide from GCRs. Some
large SEP events, however, include solar cosmic rays with energies in excess of
10\,GeV; these are efficient producers of cosmogenic
radionuclides. Their relative contribution to an annual GCR production
is small \citep{2006GeoRL..3308107U}. This is particularly true for $^{14}$C and
$^{10}$Be ($^{36}$Cl is more sensitive to lower energies and
is therefore a promising candidate to study strong SEP’s, but as
$^{36}$Cl is produced by spallation of the relatively rare $^{40}$Ar
this reduces the temporal
resolution for standard-sized ice cores or requires considerably larger ice samples to measure
it with the required signal-to-noise contrast).

SEP’s recorded by particle detectors during the past 50 years show a
range of fluences and
spectral steepness. Very large SEP events from activity near solar central
meridian typically have higher fluences, yet steeper spectra, which
makes them deficient in particles exceeding 1\,GeV.  Very large SEP events
from near the west limb of the Sun typically have lower fluences but
flatter spectra, favoring particle energies in excess
of 1\,GeV \citep{1975SoPh...41..189V}.  

All of the above effects needs to be factored in when translating
radionuclide concentrations to SEP fluences. This leads to substantial
differences in estimates. For example, 
there are three $^{14}$C production models that differ
markedly in their estimates of SEP fluences. The first estimate
was made by \cite{lingenfelter+ramaty1970}, based on an empirical
parametrization of early measurements of neutron fluxes in the Earth's
atmosphere. It predicts that the average SEP production rate for 
a year is $\sim 6$\%\ of the GCR annual
rate, and that the event of 1956/02/23 (the largest observed ground
level enhancement --~GLE~-- by neutron monitors
\citep[e.g.,][]{rishbeth09}, with a very hard spectrum) would give 1/3rd of the overall
annual $^{14}$C production. It is important to note that 
\cite{lingenfelter+ramaty1970} made the rather extreme assumption
that the magnetic shielding is reduced by a factor of 5 during large
solar storms, which leads to  a high $^{14}$C production rate.

The next estimate was based on a semi-empirical model by D. Lal
\citep{castagnoli80,lal88}, who adjusted numerical calculations to fit
empirical data. That model yields an average production for $^{14}$C
by SEP events being less than 1\%\ on average, while the event of
1956/02/23 would yield only several percent of the annual radiocarbon
production \citep{2006GeoRL..3308107U}. 

A more recent model based on an
extensive Monte-Carlo simulation of the atmospheric particle cascade
\citep{1999JGR...10412099M,2009JGRD..11411103M} suggests that SEPs contribute, in an 
average  year, only 0.03\% to production of $^{14}$C. This very small value
may be caused by the neglect of the atmospheric cascade (and thus
neutron capture channel of $^{14}$C) in their model
\citep[cf.][]{masarik95}. The most recent Monte-Carlo model
\citep{kovaltsovetal2012} suggest that the
average contribution of SEPs into the global $^{14}$C is about 0.2\%.

Thus, the model predictions differ by more than two orders of
magnitude. For the purpose of the present study, we opt for the most
conservative upper limit currently published, based on the work by
\cite{1999JGR...10412099M,2009JGRD..11411103M}: to achieve this, we
took the data by \cite{lingenfelter+hudson80}, and shifted them upward in
fluence by two orders of magnitude. These data, shown in
Fig.~\ref{fig:2b}, support a substantial drop below any power law that
can be fit to the satellite observations for events with cumulative
frequencies larger than once per century. The $^{14}$C data are
clearly more restrictive in that respect than the lunar rock data, in
that they lie further below the trend found in the directly observable
fluence range.

Calculations by \citet{2006GeoRL..3308107U} and
\cite{2007JGRA..11210106W}, based on the measured spectra of the
largest SEP in the past 50 years, predicted undetectable effects for
$^{10}$Be, $^{14}$C and $^{37}$Cl assuming global atmospheric mixing,
or a barely detectable effect if $^{10}$Be is dominated by polar
production.  This is a consequence of (a) the large amplitude of the
GCR modulation by the 
sunspot cycle that dominates the contributions by SEPs, and (b) the
high standard deviation ($\sim$15\%) of annual $^{10}$Be data. Larger
SEP events may have happened in the past, however, and increased
sample sizes, multiple cores extending back thousands of years, and
better understanding of the heliospheric variability on GCR fluxes may
make it possible to use radionuclides to inform us on the SEP fluence
frequency distribution as shown in Fig.~\ref{fig:2b} for frequencies
below once per few decades. But achieving such results requires
considerable analysis, well beyond what is feasible in the present
study.
 
\section{Nitrate concentrations in ice, and the possible link to solar particle events}\label{sec:nitrates}
When solar energetic particles impact the Earth's atmosphere they
cause ionization in the polar regions that results in production of NO
\citep{1990JGR....95.7417J,1993GeoRL..20..459J,jackman+etal2008}. 
The NO is interconverted to other
odd nitrogen species, and some of it, at whatever altitude it is
produced, should ultimately end up deposited in
snowfalls as nitrate (in aerosol or scavenged from gaseous HNO$_3$),
if not destroyed prior to that by chemical interactions at mesospheric
and higher layers. Given that SEPs most readily enter the Earth's
atmosphere at high geomagnetic latitudes, and because long-lived ice
is readily found at high geographic latitudes, it is logical to seek a
nitrate signal in polar ice cores.

\cite{1990SoPh..127..333D} reported on the analysis of two ice cores
from Windless Bight in Antarctica in which they measured the nitrate
concentration going back to about 1905\,AD.  The Antarctic record was
later supplemented by a core from Summit in Greenland (GISP2 H core)
which was measured at a sampling density of 10 to 20 samples per year
(for samples of 1.5\,cm in thickness) extending back to
1561\,AD.

The Summit core contains spikes in the nitrate concentration that are
superposed on a regular seasonal cycle. These spikes are often just 1
sample wide but occasionally 2-3 samples wide, i.e., occur in a period
that could range from a single snowfall up to about 3 months.  The
core contains a continuous spectrum of spikes, from many small ones,
to over one hundred large to very large ones: the largest spikes are
about a factor 5 larger than the typical seasonal cycle amplitude.
Dating of these spikes is achieved by counting annual layers in the
cores, supplemented by identification of deposits associated with
strong known volcanic eruptions. With that information, the year
should be accurately known near the volcanic markers (32 over the
430-y record), but might deviate by a year or two away from such
markers. 

The coincidence of some of these spikes with known space-weather events
suggested that at least the strongest among them might originate from 
SEP events.  In particular, the strongest (integrated) peak in the Summit-core
record was dated to within a few weeks of the 1859 Carrington event,
one of the largest known solar flares and associated CME sequences to
impact geospace \citep{1994kans.rept.....D,1993ICRC....3..846S,2001JGR...10621585M,2003JGRA..108.1268T,2004SoPh..224..407C,2006AdSpR..38..232S}.

The timing and sharpness of the nitrate spikes is
problematic if nitrate is indeed associated with
SEPs, because it is difficult to transport nitrates from above the
tropopause quickly enough into tropospheric snowfalls within a matter
of at most a few weeks. Furthermore, one would not expect nitrate
produced in the middle atmosphere to be deposited over such a short
time period, nor would one expect troposheric enhancements by a large
factor, as observed in the spikes. This could be resolved if the snow is
actually recording a tropospheric, rather than a stratospheric,
production of nitrate.  Alternatively, if the SEP event is having its
effect higher in the atmosphere, it may be accompanied by one of the
rare (especially in Antarctica) sudden stratospheric warming events
which could transport material downwards relatively rapidly, perhaps
allowing a response within a month or two. However, the coincidence of
two such rare events would be unusual.  Otherwise, one would expect a
transport time of order 6 months, and thus no sharp signal in the ice
core chemical patterns. 

Strong nitrate spikes may be caused by terrestrial events or by
depositional processes. It has been well documented that nitrate
spikes associated with enhanced ammonium concentrations are an
indication of biomass burning, and these are seen in Greenland ice
cores, including those from the Summit regions where the H core was
taken
\citep[e.g.,][]{legrand+etal1992,whitlow+etal1994,fuhrer+legrand1997}.
Spikes can be induced by changes in scavenging efficiency owing to,
for instance, changes in the degree of riming (the inclusion of supercooled droplets as snow crystals grow). More specifically,
spikes in nitrate deposition are induced by conversion of nitric acid
to aerosol through association with either sea salt (for coastal
Antarctica) or ammonia (for central Greenland) leading to deposition
of the associated aerosol \citep{wolff+etal2008}.

In the work leading up to this manuscript, \cite{wolff+etal2012} assembled information on a total of 14 ice
cores with high time resolution from both arctic and antarctic
regions, at various geomagnetic latitudes. They found that apart from
the Summit GISP2\,H ice core, no nitrate signatures were found in the
ice dated to 1859.  Several nitrate spikes of a similar nature were
found in all the Greenland cores, including one from the Summit site.
However, all such spikes including one dated to 1863 (the nearest large
spike to 1859 in the later records), were associated with an ammonium
spike.  In the cores where other components were measured, black
carbon and vanillic acid (diagnostic of combustion plumes in general,
and wood burning, respectively) were found in each large
spike between 1840 and 1880.  None of these components were measured
in the H core, so \cite{wolff+etal2012}  cannot conclusively identify the origin of the
peak labelled as 1859, but do conclude that it is inevitable that most
nitrate spikes in all Greenland cores are of biomass burning origin.
While it may be possible to isolate very large events that are not of
such origin, \cite{wolff+etal2012}  conclude that even the 1859 event was
not large enough to give a signal in most ice cores.  It is
unfortunately apparent that the statistics of nitrate cannot provide
the statistics of SEP
events so that this potential proxy for SEP events
prior to the mid 20th Century can, at present, neither be used to
estimate the frequency spectrum of SEP events nor to set unambiguous upper
limits to a possible historical maximum for such events.  In view of
this, the nitrate data shown in figures by \cite{2001JGR...10621585M} and
\cite{usoskin_2008} are not included in Fig~\ref{fig:2b}.

\section{Flare energies}\label{sec:flares}
\subsection{Solar flares}\label{sec:solarflares}
In order to compare the occurrence frequencies of solar and stellar
flares as a function of their energy, the diversity of available
measurements needs to be transformed to a single unified scale. Here,
we attempt to rescale the observations to bolometric
fluences, based on available approximate conversions. 

\begin{figure}[t]
\noindent\includegraphics[width=8.4cm]{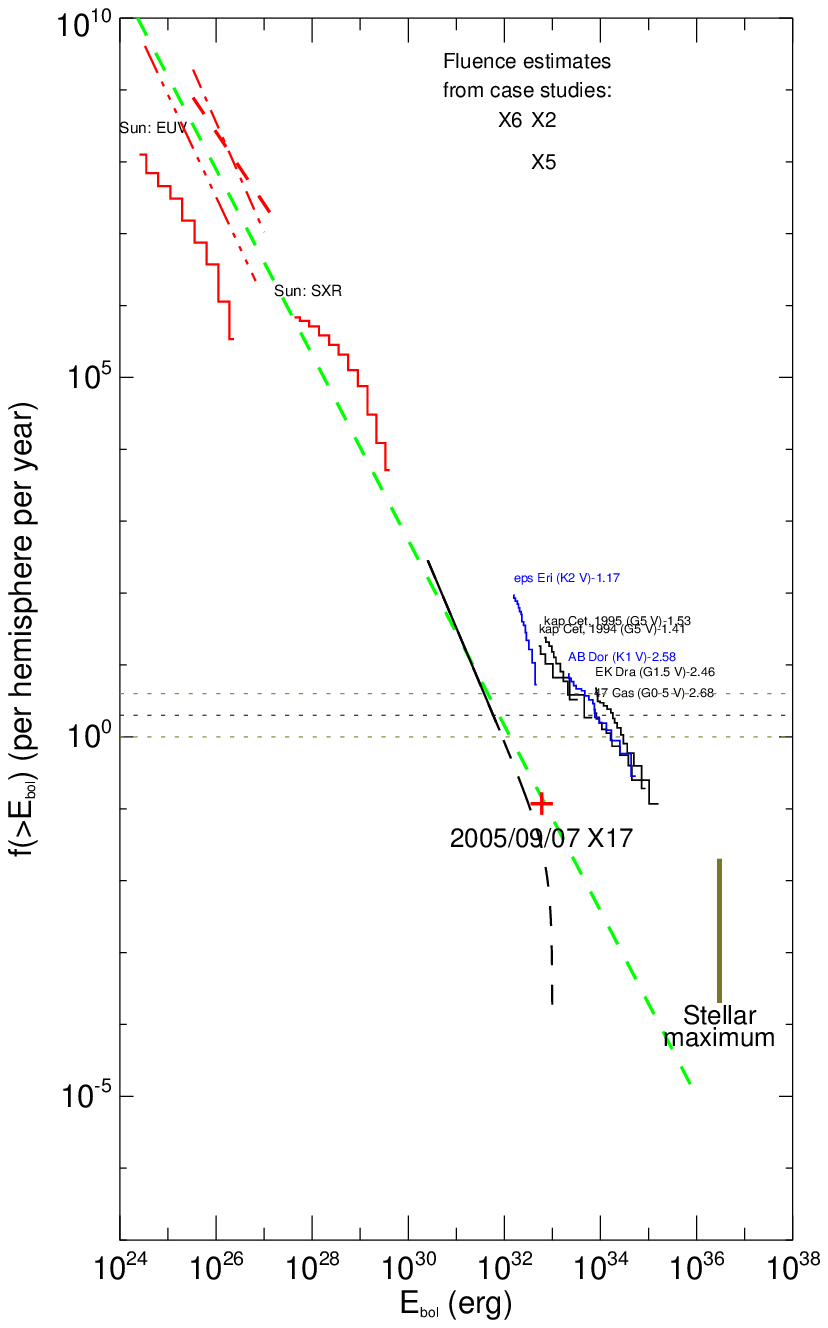}
\label{fig:2a}
\caption{\em Downward-cumulative frequency distribution for
  bolometric flare fluences, $E_{\rm bol}$. Quiet-Sun EUV microflares:
  solid histogram: \citet{2000ApJ...535.1047A}; dash-dotted:
  \citet{krucker+benz99}; dashed-triple-dotted:
  \citet{parnell+jupp2000}; dashed:
  \citet{2002ApJ...568..413B}. Lower red histogram: active-region soft X-ray flare data
  from \citet{shimizu95,shimizu1997}. The
  dashed black curve shows Eq.~(\ref{eq:solarflarefluencespectrum})
  for $F_{\rm max}= 10^{33}$\,ergs; the empirical range of 
  a power-law approximation is shown solid.
  A red cross marks the fluence of the
  2005/09/07 X17 flare at an equivalent cumulative frequency for it to
  be the 5th largest flare since 1976 (see
  http://www.spaceweather.com/solarflares/topflares.html).  Histograms
  for $E_{\rm bol}\ga 10^{32}$\,ergs are for active Sun-like
  main-sequence stars (black or blue for spectral type G or K),
  scaled as in Eq.~(\ref{eq:stellar2solar}). The vertical bar marks
  the largest flare fluence in Sun-like stars in Kepler
  observations. The green dashed power law, with index $-2.3$,
  approximately connects the solar data.  The central dotted
  line shows the frequency at which CME opening angles reach $2\pi$
  radians (with its uncertainty range).}
\end{figure}
For solar
flares, characteristically $\approx 70$\%\ of a flare's total
radiative energy is emitted at visible wavelengths 
(characterized by a
blackbody temperature of approximately $9000$\,K; see, e.g.,
\cite{2004GeoRL..3110802W,2007ApJ...656.1187F}; that value
is also found for stellar flares, see, e.g.,
\cite{1992ApJS...78..565H}). This
can be used to begin the comparison of energy scales to the GOES flare
classification scale, Fig.~\ref{fig:2a} shows (at the top) the total
energy estimates for three well-studied solar flares \citep[of classes
X2, X5, and X6;][]{aschwanden+alexander2001,benz2008}, assuming that
these X-ray and EUV estimates are complemented by another 70\%\ of the
total energy to make the bolometric fluence as described above. These
flares suggest that the average X4.3 flare would have a bolometric
fluence of $4.9\times 10^{32}$\,ergs.

\cite{2006JGRA..11110S14W}
provide excess total solar irradiance (TSI, i.e., the flare fluence) estimates for four very energetic flares. Two of those, which
occur well away from the solar limb, are X17 and X10 flares with
fluences of $6.0\times 10^{32}$\,ergs and $2.6\times 10^{32}$\,ergs,
respectively.  The average of these fluences is about that estimated
above for an average X4.3 flare, beginning to illustrate the
uncertainties in the conversion process to a common bolometric scale. 

On average,
only $\approx 0.6\pm 0.1$\%\ of the total photon energy is emitted in
the GOES $1-8$\,\AA\ channel that is used to classify flares by their
peak intensity\citep{emslie+etal2005,2011A&A...530A..84K}. These numbers, derived from
composite observations of C-class to X-class flares \citep{2011A&A...530A..84K}
require an average multiplier of $\approx 160\pm 30$ to convert a fluence
derived from the GOES $1-8$\,\AA\ passband to a bolometric
fluence. For comparison, direct total solar irradiance measurements for four large
flares \citep{2006JGRA..11110S14W} suggest multipliers of $\approx
49-162$, with values of $126$ and $162$ for the \referee{two flares (X17 on 28
Oct.\ 2003 and X10 on 29 Oct. 2003) well away from
the solar limb, roughly consistent with the abovementioned average conversion.} 

\referee{In his study, \citet{2011A&A...530A..84K} differentiates
  flares into four groups (C4-M2.8, M2.8-M6.4, M6.4-X1.3, and
  X1.3-X17) and uses a superposed epoch analysis for all flares
  within these subgroups to derive conversion factors from GOES
  $1-8$\AA\ fluences to SOHO/VIRGO bolometric (or total solar
  irradiance, TSI) fluences. The conversion
  factors for the four subgroups are $330\pm 130$, $220\pm 80$,
  $140\pm 60$, and $90 \pm 10$. These results show a decrease in the
  conversion factor with increasing flare magnitude, for TSI fluences from
  $0.36\times 10^{31}$\,ergs to $5.9\times 10^{31}$\,ergs. The
  conversion factor for the group of X-class flares is some 35\%\
  lower than those described above, which may be a consequence of
  differences in samples or, for example, be influenced by positions
  on the solar disk. In the remainder of this study we use a power-law
approximation of the conversion from $1-8$\AA\ GOES fluence to TSI fluence
provided by \citet{2011A&A...530A..84K},}
\begin{equation}\label{eq:goestsi}
F_{\rm TSI}=2.4\times 10^{12} F_{\rm GOES}^{0.65\pm 0.05},
\end{equation}
\referee{although 
we give preference to direct bolometric
fluences for those large flares for which these where published.}

Other estimates of bolometric flare energies are available in the
literature, but generally these are subject to assumptions that may
cause these estimates to be significantly different from direct
observations
of the total solar irradiance (TSI), so they are excluded here (for example, the energy in
$>20$\,keV electrons in the X28+ flare on 2003/11/04 has been
estimated to be of order $\approx 1.3\times 10^{34}$\,erg
\citep{2005A&A...433.1133K}, but see \cite{2009SoPh..258..141T} 
for an alternative view of the implications of these observations).

GOES observations revealed a soft X-ray
($1-8$\,\AA) flare fluence distribution \citep{2002A&A...382.1070V}
that transforms to a downward-cumulative distribution function for
bolometric fluences\referee{(applying Eq.~\ref{eq:goestsi}))  of
\begin{equation}\label{eq:solarflarefluencespectrum}
N^\ast_f(F_b)  =  9.2\times 10^{33}\,\left ({1\over F_b^{1.03 \pm 0.09}} - {1\over F_{\rm max}^{1.03 \pm 0.09}}  \right ),
\end{equation}
where $F_{\rm max}$ is a possible cutoff fluence beyond which
no flares occur (discussed below). In deriving this distribution,
\citet{2002A&A...382.1070V} did not correct for the background X-ray
emission beneath the flare emission; such a correction would be
important for relatively small flares, but as we focus on M-class
flares and larger (with the above power law approximation valid only
starting at mid-C class flares), the effects are limited and ignored below.}

The largest observed flare saturated the GOES detector and was
estimated to peak at X28, not much above the X10 and X17 flares
discussed above. Hence, for the purpose of illustration (and arguments
below) we assume a value of $F_{\rm max} =10^{32}$\,ergs, about twice the abovementioned
average flare fluence for the X10 and X17 flares as a lowest likely
upper limit to flare fluences. This would approximately correspond to
an X25 flare using the scaling that GOES soft X-ray fluence $F$ and
the GOES flare class (the peak brightness) $B$ are related through
\begin{equation}\label{eq:bf}
F\propto B^{1.10}
\end{equation}
\citep{2002A&A...382.1070V}.  
Fig.~\ref{fig:2a} shows the above distribution for $F_{\rm
  max}=10^{33}$\,ergs  as a dashed black curve.  

This distribution is based on 8400 flares from 1997 through 2000 for
which GOES $1-8$\AA\ fluences were specified; the power law holds for
flares with a range of $1-8$\,\AA\ GOES fluences from $\approx 6\times
10^{27}$ to $\approx 10^{30}$\,ergs. \referee{The normalization of the
  annual frequency distribution in
  Eq.~(\ref{eq:solarflarefluencespectrum}) and Fig.~\ref{fig:2a} for
  an average over a full solar cycle is achieved by setting the
  frequency for an M1-class flare to the average frequency of 140 per
  year for flares of M1 or larger over the period of cycle 23, from
  1996/01/01 to 2007/01/01, and taking a value of $4 \times
  10^{30}$\,ergs as the bolometric fluence for a characteristic M1
  flare from \citet{2011A&A...530A..84K}.}

For flares below GOES class C, the determination of the flare
frequency distribution from the disk-integrated GOES signal becomes
increasingly ambiguous for less-energetic events. As one goes down the
energy scale, the signal from individual flares sinks into the
background soft X-ray luminosity. Moreover, the flare photon spectrum
weakens in X rays and strengthens in the extreme ultraviolet (EUV) for
flares of decreasing magnitude. For the study of less energetic
events, spatially-resolved X-ray or EUV imaging is more appropriate.
One such study used Yohkoh soft X-ray images
\citep{shimizu95,shimizu1997} to estimate flare energies from imaging
observations of an active region and its immediate surroundings in a
field of view of 5\,arcmin square. These are observations of only a
single moderately large active region. The histogram in
Fig.~\ref{fig:2a} shows the results of this study assuming that
averaged over a solar cycle 3 such regions exist on the disk, and
using the estimate that 70\%\ of the energy is emitted at visible
rather than X/EUV wavelengths.

For even smaller flares, several energy fluence distributions are
available based on either the SOHO/EIT or TRACE EUV observations
\citep{krucker+benz99,parnell+jupp2000,2000ApJ...535.1047A,2002ApJ...568..413B}. In
order to convert these energies to estimated bolometric fluences we
use the finding that approximately 15\%\ of the event energy is
emitted in the coronal EUV, as derived for larger flares
\citep{2011A&A...530A..84K}, although this has not been verified for
the smaller events observed in the EUV only.  The differences between
the four distributions shown are related to different algorithms for
flare characterization and to assumptions about the geometrical extent
of the observed events along the line of sight
\citep{2000ApJ...535.1047A} .

\referee{For the remainder of this study we are primarily interested
  in the largest flares, but we point out that it is intriguing that
  these solar flare distributions align relatively well, within the
  substantial uncertaintites in energy conversions and from the
  perspective of a log-log diagram. A rough power-law approximation
  (shown by the green dashed line in Fig.~\ref{fig:2a}, is given by
\begin{equation}\label{eq:solarflaredist}
N(E_f) {\rm d}E_f \propto E_f^{\alpha_f} {\rm d}E_f,
\end{equation}
with $alpha_f = -2.3 \pm 0.2$, with an estimated uncertainty 
that is largely associated with the uncertainties in the conversions from
X-ray and EUV fluences to bolometric fluences for microflares to large solar flares.}

\subsection{Stellar flares}\label{sec:stellarflares}
Solar flares are a manifestation of the Sun’s magnetic field, and that
field is believed to arise from the interaction of convection with
rotation, especially differential rotation: the dynamo mechanism.
Other stars with convective envelopes (G, K, and M spectral types)
also show magnetic activity, including flares.  Here we will discuss
G- and K-type stars because they are most similar to the Sun.  Stellar
flares cannot be resolved spatially and so we can detect only
energetic events that produce sufficient contrast against the visible
photosphere or the X-ray/radio corona.  In addition, stellar
observations often are available for only a limited wavelength range
and so it is difficult to gain a full bolometric view of an event.
Detectable high-energy flares have been seen on rapidly-rotating GKM
stars because the rotation enhances the magnetic field.  Single GKM
main sequence stars lose angular momentum with age and so the flaring
stars are either very young (up to $\sim 100$\,Myr old), or they are in
close binaries where tidal interaction causes spin-up of an older
star; these latter systems are known as BY Dra binaries (main
sequence) or RS CVn binaries (evolved).  Stellar flares have been
reported with X-ray or EUV energies as low as $\sim 10^{28}$ ergs
\citep{2002ApJ...580L..73G}; this corresponds to a bolometric fluence $\sim
3-5$ times higher. Most reports on stellar flaring report fluences
much larger than that simply
because of the difficulty of detecting small flares against the bright
background of the overall corona or photosphere.

More energetic flares have harder emission, and so the passband used
biases the detection threshold.  Observations in the shorter-wavelength X-rays tend to
favor the largest flares, making the energy distribution appear less
steep than it really is.  This has been seen explicitly in BeppoSAX
observations with soft (about 0.2-2 keV) and hard ($>1$\,keV)
channels.  Simultaneous observations on the same flares made in both
bands led to a slope $\alpha = 2.4 \pm 0.2$ for the soft channel and
$\alpha = 2.0 - 2.2$ for the hard channel \citep{guedel+etal2003}. For
this reason, it is preferable to search for flares and coronal
radiation in either soft X-rays or the EUV.

Figure~\ref{fig:2a} shows stellar flare data for five G and K main
sequence stars \citep[from][]{audard+etal2000a} in soft X-ray and EUV
bands ($0.01 - 10$\,keV, or $1.2 - 1200$\,\AA), scaled to approximate
bolometric fluences by assuming the same ratio between bolometric and
coronal fluences holds as for solar EUV observations (see
Sect.~\ref{sec:solarflares}), i.e., that about 30\% of an event’s
energy is emitted in the EUV \citep[excluding M-type stars which show
comparable behavior but are far from solar in their basic properties;
data from][]{audard+etal2000a}.  Little is known in the literature
about the ratio of coronal to bolometric brightness during flares. One
example of a large flare on an ultracool M8 dwarf star
\citep{2006A&A...460L..35S} showed comparable energies in the visible
and soft X-ray passbands in which the flaring star was observed, in
acceptable agreement with our assumption for purpose of comparison of
solar and stellar data in Fig.~\ref{fig:2b}. In the absence of further
information, we make the simplest assumption, namely, that solar and
stellar flare energies are, to first order, similarly distributed over
the electromagnetic spectrum.

\referee{The five G- and K-type stars for which
  \citet{audard+etal2000a} determined the flare frequency
  distributions are highly active and rotate much more rapidly than
  the Sun. The most active among these stars exhibit flaring at
  energies of $10^{33}$\,erg several times per day. The studies by
  \citet{osten+brown1999} and by \citet{audard+etal2000a} revealed
  that the frequency of flaring in these stars increases nearly
  proportionally to the background stellar X-ray luminosity which
  spans a range of a factor of $10^4$ in their sample (their
  Fig.~4). \citet{audard+etal2000a}, for example, find a power-law
  index of $0.95 \pm 0.10$ for the scalings between cumulative flare
  frequencies and coronal X-ray luminosity. For the comparison shown
  in Fig.~\ref{fig:2a} we assumed a purely linear dependence, so that
  the observed cumulative distribution of flare energies, $N_{\rm
    obs}(>E|L_{\ast,X})$, for a star with X-ray luminosity $L_{\ast,X}
  $ transforms to the distribution $N^\odot(>E)$ scaled to the solar
  X-ray luminosity, $L_{\odot,X}$ through
\begin{equation}\label{eq:stellar2solar}
N^\odot(>E) = \left( {L_{\odot,X}  \over L_{\ast,X} }\right) N_{\rm obs}(>E).
\end{equation}
This scaling shifts the stellar distributions downward in the diagram
towards the solar distribution, while essentially collapsing them onto
each other. In this scaling, we used  an estimated average
coronal luminosity for the Sun of $L_{\odot,X} =4.3\times
10^{27}$\,erg\,s$^{-1}$, which is an average over the solar cycle
for the $0.1-2.4$\,keV bandpass \citep{judge+etal2003f}.}

\referee{The comparison of the solar and scaled stellar frequency
  distributions for flare energies in Fig.~\ref{fig:2a} shows that the
  frequency distribution for large solar flares lies substantially
  below the scaled stellar data. From this we conclude that the data
  on active stars cannot be used to infer the probabilities of
  solar flares of high energies that may or may not occur at frequencies 
  below once per few decades.}

In the decade following the work by \citet{audard+etal2000a},
energetic flares have been seen in G stars in the white-light bandpass
($\sim$ 4000-9000\AA) of the {\it Kepler} mission.  About 0.5\%\ of
the brightest G stars exhibit flares with fluences of $\sim 10^{34}$
up to $\sim 10^{37}$ ergs
\citep{2011AJ....141...20B,walkowicz+etal2010}. These are energies in
the optical bandpass and so are lower limits since some energy is
emitted at other wavelengths.  Flares exceeding $10^{37}$ ergs have
not yet been seen.

The G stars that show flares in the {\it Kepler} data most often show
multiple flares, with some flaring about every other day.  Most {\it
  Kepler} data is sampled every 30\,minutes and so only very energetic,
long-lived flares are reliably detected.  However, a subset of stars
is observed every minute and a few flaring stars have been so
observed.  In those cases it is possible to fully resolve the rise of
a flare (with a time scale of $\sim 10$\,min) and its decays
(on time scales of hours), with secondary events during the decay
\citep{soderblom+etal2012}.  The very large sample size of {\it Kepler}
(some 150,000 stars) corresponds to an effective monitoring time for a
single average G star of
$\sim 400,000$ years, far beyond anything previously done.

Another very energetic flare was reported for II\,Peg
(K2IV+dM) at $10^{37}$\,ergs \citep{2007ApJ...654.1052O,2008ApJ...688.1315E}.   
Other reports of very large flares include
\cite{1996A&A...311..211K} 
who  observed a flare from the binary CF\,Tuc with radiated energy in the ROSAT bandpass of $1.4\times 10^{37}$\,erg, and 
\cite{1997A&A...328..565E} 
who reported on a large flare from HU\,Vir which had a radiated energy of $7.7\times 30^{36}$\,ergs in the same bandpass; both of these targets are active binaries. One
extreme value is $10^{38}$\,ergs reported by \cite{schaefer+etal2000}, 
but it remains to be seen if the
source of the flare was correctly identified. 

From the above, it appears that it is highly unlikely that any flare
would exceed $10^{37}$\,ergs on a Sun-like star in any phase during
its evolution once it has comfortably settled on the stellar main
sequence. But that leaves a factor of $\sim 10^4$ between the largest
observed solar flare and the largest possible for a Sun-like star. Are
there other empirical constraints that help us narrow that gap?

\section{Mapping SEP fluences to flare energies}\label{sec:streaming}
Figure~\ref{fig:2b} shows that the slope for flare electromagnetic fluences 
(Fig.~\ref{fig:2a}) is very different from the slope seen for SEP
fluences: the power-law exponent ($\mu$) in the
frequency distributions of power-law form
\begin{equation}
dN/dx = Ax^{-\mu}  
\end{equation}
is smaller for SEP fluences --~$\mu \approx 1.1-1.3$ in
Fig.~\ref{fig:2b} below a fluence of about $5\times
10^{9}$\,cm$^{-2}$~-- than it is for flare electromagnetic emissions
($\mu =\alpha_f \approx 2.3$, see Eq.~[\ref{eq:solarflaredist}]),
while the SEP event fluence spectrum turns to a significantly steeper
spectrum above $\sim 5\times 10^{9}$\,cm$^{-2}$ \citep[see
also][]{1975SoPh...41..189V}.  Several effects may be at play here: 1)
SEP spectral distributions may depend on event energy (which could
include a dependence on the partitioning between flare radiative and
CME bulk-kinetic energies), 2) background corrections, 3) effects of
compound events involving two or more CME/shocks on SEP size
distribution, and 4) particle propagation effects in the heliosphere.

Before considering the effects of any of the above potential
processes,
we should allow for a geometrical effect that must play a role:
dilution of the fluence over an opening angle into the heliosphere,
and -~related to that~- the possibility that the SEP event misses the Earth
altogether: as SEPs propagate into the heliosphere over a solid angle
less than $2\pi$ we certainly need to correct for the probabibility
that SEP events may not hit Earth and thus not be recorded, while if 
that opening angle depends on the energy of the event, then the SEP
fluence needs to be corrected for the change of opening angle with
total event energy. We can make the following plausible quantitative argument:

Following the reasoning by \cite{schrijver2011}, although in part in the
opposite direction, we start from the observation that the frequency
distribution of particle fluences can be approximated by
a power law up to about $5\times 10^{9}$\,cm$^{-2}$: 
\begin{equation}\label{eq:fluencedist}
N_{\rm p} {\rm d}F_{\rm p} \propto F_{\rm p}^{-\delta} {\rm d}  F_{\rm p} .
\end{equation}

Let us assume that
the particles are emitted from their source region at or near the Sun
into a solid angle $\Omega$ that is a function of the total energy
$E_{\rm f}$ of the event, here chosen to be approximated by:
\begin{equation}\label{eq:eruptangle}
\Omega \propto E_{\rm f}^\gamma .
\end{equation}
The value of $\gamma$ can be estimated by comparing the flare 
energy distribution in Eq.~(\ref{eq:solarflaredist}) with a  distribution
of opening angles, $a$ (in degrees), for eruptions from 
small fibril eruptions to large CMEs \citep[summarized by][]{schrijver2010}
\begin{equation}\label{eq:angledist}
N_a {\rm d}a = b\, a^{-\beta} {\rm d a},
\end{equation}
with $\beta=2.0 \pm 0.3$ (with $b\approx 1.1$ for $\beta =2$). 

For given $a$ (expressed in radians), the corresponding fractional solid angle
is given by 
\begin{equation}\label{eq:openingangle}
{\Omega \over 4\pi} = {1 \over 2} ( 1- \cos{a}) \approx {1 \over 4} a^2,
\end{equation}
where the righthand exression holds for sufficiently small $a$. Using
that expression with Eqs.~(\ref{eq:eruptangle}),
and~(\ref{eq:angledist}), we find
\begin{equation}\label{eq:solidangledist}
N_a {\rm d}a \propto a^{{2 \over \gamma}(1-\alpha_f)-1} {\rm d}a.
\end{equation}
With Eq.~(\ref{eq:angledist}) we find 
$\gamma = 2(\alpha_f-1)/(\beta-1) \approx 1.9 \pm 0.6$.

If the particle fluence at Earth, $F_p$, is a fixed fraction $f$ of
$E_{\rm f}$, diluted by expanding over a solid angle $\Omega$, then
with Eq.~(\ref{eq:eruptangle}),
\begin{equation}\label{eq:}
F_p \propto f {E_{\rm f} \over \Omega} \propto E_{\rm f}^{1-\gamma},
\end{equation}
transforming Eq.~(\ref{eq:solarflaredist}) to read
\begin{equation}\label{eq:fluencedist2}
N_{\rm p} {\rm d}F_{\rm p} \propto F_{\rm p}^{{1-\alpha_f \over 1- \gamma}} {\rm d}F_{\rm p}.
\end{equation}

As in this model experiment SEPs are assumed to be emitted within a
solid angle $\Omega$, only a fraction
\begin{equation}\label{eq:probability}
p = {\Omega \over 4\pi} \propto E_{\rm f}^\gamma 
\end{equation}
of the total number of events can be detected near Earth. Hence, to
derive the  SEP fluence 
distribution from the flare energy fluences,
Eq.~(\ref{eq:fluencedist2}) has to be multiplied by $p$:
\begin{equation}\label{eq:fluencedist3}
N_p {\rm d}F_{\rm p} \propto F_{\rm p}^{{1-\alpha_f +\gamma \over 1- \gamma}}
{\rm d}F_{\rm p}\equiv F_{\rm p}^{-\epsilon}{\rm d}F_{\rm p}.
\end{equation}
With the values of the exponents above, we find $\epsilon =
(\alpha_f-1)(\beta-3)/(\beta-2\alpha_f+1)=-0.8\pm0.2$, consistent with
the observations provided that we limit the comparison to events for which
the SEPs are spread over a solid angle small compared to $2\pi$
steradians. 

An opening angle of 180$^\circ$ is reached for an event frequency of
approximately twice per year \citep{schrijver2010}, with an
uncertainty of at least a factor of two. That range, shown by dotted
horizontal lines in the panels of Fig.~\ref{fig:2a} and~\ref{fig:2b}, lies just above
the frequency where the SEP event fluence frequency distribution bends
downward, suggesting that geometrical considerations may be a dominant
effect in changing the slope of flare to SEP fluences at least around
the range labeled '(i)' in Fig.~\ref{fig:2b}, but not for energies at or
above the value labeled '(ii)'.  In other words, the
segment of the observed SEP fluence distribution function labeled
'(i)' in Fig.~\ref{fig:2b} likely needs to be steepened to accommodate
the above geometrical effects, and this steepening appears to bring it
in line with the slope found for flare bolometric fluences, i.e., with
the green dashed line.  Therefore, the break in the SEP fluence
spectrum above the downward kink could reflect a
limit on the spreading of the SEPs in angle.  We note, however, that we
have insufficient information on the angular width distribution of SEP
events in general: observations put many of these opening angles for
impulsive events on a gaussian-like distribution with $\sigma=15^\circ
- 20^\circ$ \citep{1999SSRv...90..413R} whereas recent STEREO
observations have shown events with opening angles up to $136^\circ$
\citep{wiedenbecketal2011}. Nevertheless, this argument offers a
plausible origin to the kink in the frequency distribution in
Fig.~\ref{fig:2b} so that we cannot assume that kink is
indicative of a change in the behavior of solar flare fluences for the
largest flares. 

\section{Conversion of magnetic energy to power flares}\label{sec:conversion}
Having established that currently available flare statistics on
Sun-like stars are not directly applicable to the present-day Sun
owing to the difference in mean activity level, and that lunar and
terrestrial records leave a substantial range of uncertainty on the
largest solar events, we explore one further avenue. The energy
released in large solar coronal storms is ultimately extracted from
the electromagnetic field in the solar atmosphere. Because that energy
is associated with the surface magnetic field, including its sunspots,
some constraint may be derivable from sunspot sightings.

One element of this argument is the observation that mature active
regions -~within a bounding perimeter including spots, pores, knots,
and faculae~- are characterized by a remarkably similar flux density,
$B_0=\langle B \rangle$ of about $100$\,Mx/cm$^2$ to $150$\,Mx/cm$^2$
\citep{ref254} regardless of region size. This allows us to perform an
order of magnitude scaling between the energy available for flaring in
the magnetic field above an active region and the flux that this
region contains.

If we assume that a fraction of $f=0.01-0.5$ of the magnetic
energy density in a volume with a characteristic mean field strength
of $B_0$ can be converted into what eventually is radiated from the
flare site \citep[e.g.][]{metcalf+etal2005,schrijver+etal2007}, the typical dimension $d_0$ and magnetic flux $\Phi_0=B_0
d_0^2$ in such a flaring region are given by
\begin{equation}
d_0=\left ( {4\pi E_f/f \over B_0^2} \right)^{1/3} \,\,;\,\, \Phi_o=B_0 d_0^2
= {(4\pi E_f/f)^{2/3}\over B_o^{1/3}}. 
\end{equation}
For a very large flare with energy $E_f=10^{37}$\,ergs, we find
$d_0\approx (2-7)R_\odot$ and $\Phi_0\approx (10-80)\,10^{24}$\,Mx,
even using $B_0=300$\,G; \referee{to illustrate the magnitude of the problem, 
we chose a value of $B_0$ for this estimate that is, in fact, $2-3$ times higher than
characteristic of solar regions \citep{ref128,ref254}. Even with an average magnetic flux
density substantially above what the present-day Sun shows us,} 
the flaring region simply would not fit on the Sun.  For
flares with $E_f=10^{35}$\,ergs, $d_0\approx (0.4-1.6)R_\odot$ and
$\Phi_0\approx (0.3-4)\,10^{24}$\,Mx. Although very sizeable, and
requiring a relatively large average surface field strength, these
numbers are still compatible with the size of the Sun. Are they
compatible with the largest observed regions on the Sun?

The flux distribution for historically observed
active regions reported on by \citet{zhang+etal2010} exhibits a
marked drop below the power law for fluxes exceeding $\Phi \sim 6 \times
10^{23}$\,Mx, and they find no regions above $\Phi_{\rm max} \sim 2
\times 10^{24}$\,Mx. Historically, the largest sunspot group recorded
occurred in April of 1946, with a value of 6\,milliHemispheres
\citep{taylor1989}; for an estimated field strength of 3\,kG, that
amounts to a flux in the spot group alone of $\Phi_{\rm spots} \sim 6
\times 10^{23}$\,Mx. The total flux in this spot group was likely
larger, but perhaps within a factor of $2-3$ of that in the spots, and
thus of the same order of magnitude as the upper limit to the
distribution found by \citet{zhang+etal2010}.

Starting from that largest flux of $3\Phi_{\rm spots} \sim 1.8 \times
10^{24}$\,Mx for $B_0=100$\,G and $f=0.5$, an upper limit for flare
energies of $E_f=f\Phi^{3/2}/(4\pi B_0) \approx 10^{33}$\,ergs
results, comparable to the excess-TSI fluence reported for
well-observed X17 and X10 flares well-away from the solar limb
reported by \cite{2006JGRA..11110S14W} (see
Sect.~\ref{sec:solarflares}).

In other words, a solar flare with energy of a few times
$10^{32}$\,ergs is compatible with what we know about the largest
solar active regions.  A flare with an energy of, say, $10^{34}$\,ergs
would seem to require a spot coverage some 20 times larger than the
historically observed maximum, or 12\%\ of a hemisphere (the largest
spot coverage for the Sun as a whole reported by the Royal Greenwich
Observatory since 1874 is 0.84\%). No such records of monster spots on
the Sun have been historically reported or pre-historically recorded,
so they are likely not to have occurred over the past centuries or
even millenia. In fact, our simple scaling arguments suggest that an upper limit
of close to the largest flares observed during the past three decades is consistent with the reported
observations on the largest sunspot groups over the past few centuries.

\section{Discussion and conclusions}\label{sec:discussion}
We attempted to combine direct observational records of SEP events
associated with flares and CMEs with upper limits based on lunar rock
samples, terrestrial biosphere samples, and ice-core radionuclide
concentrations to establish a frequency distribution of approximate
particle fluences (Fig.~\ref{fig:2b}). The lunar and terrestrial
samples do constrain SEP fluences for the largest events, but only as
upper limits for fluences well beyond the historical records obtained
during the space age.  Hence, this information cannot at present be
used to significantly contribute to our knowledge of the frequency
spectrum of flare energy fluences beyond the historically observed
range that extends up to about X30.

We have had to conclude that nitrate concentrations in polar ice
deposits cannot, at present, be used to extend the direct
observational records of SEP events to a longer time base without at
least significantly more study. 

Once the multiple factors influencing the $^{10}$Be data are better
understood, it may be possible to set an upper limit that will further
constrain the event frequencies for high fluence events.
This will include establishing a calibration from $^{10}$Be
concentrations to SEP event fluences. 
Should such a calibration become available in the
future, effects of limitations on the transport of energetic particles
through the heliosphere (the ``streaming limit'' discussed in
Sect.~\ref{sec:intro}) shall need to be better understood before
the $^{10}$Be upper limit can be mapped to solar flare energy fluences.

We present an argument that the ``kink''' in the \hbox{$>10$\,MeV} SEP
fluence frequency spectrum around $5\times 10^{9}$\,cm$^{-2}$ does not
necessarily reflect a change in the flare-energy spectrum, but may in
fact be a consequence of geometrical effects related to the finite
opening angle of SEP cones. This effect causes a decrease in detection
frequency for smaller opening angles simply because events with
smaller extent are more likely to miss the Earth, combined with a
dilution of the fluence over that opening angle that affects the
particle flux density. This argument is supported by the fact that the
frequency at which the kink occurs corresponds relatively well with
the frequency for which observed opening angles of CMEs approach
$2\pi$ steradians. We therefore suggest that this kink likely does not
reflect a change in the shape of the solar flare fluence distribution,
but rather that it reflects the geometry of SEP generation and
propagation.

\referee{The combination of solar and stellar flare observations shows
  that the Sun and a sample of younger, more active stars are not
  brought into alignment for their flare-energy frequency spectra even
  if their frequencies are scaled with the average background coronal
  luminosity of the star (based on an empirical scaling derived for
  stars in a range of activities much higher than that of the Sun). We
  shall need to trace how strongly the assumptions made in the
  conversion from X-ray/EUV fluences to bolometric fluences based on
  the solar flares affect this misalignment.  But regardless of the
  outcome of that, this misalignment of solar and stellar data means
  (i) that currently available data on flares on very active stars
  cannot help us in our quest to determine frequencies of extremely
  large solar flares, and (ii) that in order for stellar data to be
  helpful in that respect, observations of stars of solar type as well
  as of roughly solar activity level are required to establish the
  X-ray/EUV properties of large stellar flares as well as their
  bolometric fluences in order to be able to enter them into a
  frequency-fluence diagram as we made here for solar flares.}

\referee{The solar flare observations can be roughly approximated by a
  power law frequency distributions as in
  Eq.~(\ref{eq:solarflaredist}). If we start from the assumption that
  flare fluences follow this power-law parent distribution function
  with index $\approx -2.3$, we can establish how likely it is that we
  have a 30-y run of observations in which no flares are seen with
  energy fluences exceeding $\approx 10^{33}$\,ergs -~at a GOES class
  of roughly around X30, subject to a calibration uncertainty of at
  least 50\%\ (see Sect.~\ref{sec:solarflares})~- if the power law
  would in fact persist up to a cutoff of the most energetic among
  stellar flares, i.e., around $10^{37}$\,ergs. We find that this would occur
  once in 10 30-y samples, which, although relatively unlikely, is not statistically
  incompatible with the observations. This does not provide us with a
  significant upper bound to solar flare energies by itself, but does
  provide a probability of at most 10\%\ for any flare exceeding the
  presently observed maximum in the next 30 years.}

\referee{We argue that flares with a magnitude well above the
  observational maximum of about $10^{33}$\,ergs are unlikely to occur, however,
  by the argument presented in Sect.~\ref{sec:conversion}. Such flares 
would}
require that much of the solar surface be covered by
strong kilo-gauss fields, exhibiting large sunspots that have not been
recorded in four centuries of direct scientific observations and in 
millenia of sunrises and sunsets viewable by anyone around the
world. For example, a flare with an energy of around $10^{34}$\,ergs
should require a spot coverage of just over 10\%\ of a solar
hemisphere, which would be readily visible even to naked-eye observers
if it occurred. Sunspot records suggest that no regions were observed
in the past four centuries that could power flares larger than those
observed in the most recent three decades.

We conclude that flare energies for the present-day Sun have either a
true upper cutoff or at least a rapid drop in frequency by several
orders of magnitude below the scaled stellar frequency spectrum for
energy fluences above about X40.  Based on the direct solar
observations and the indirect arguments presented in this study, solar
flares with energy fluences above about X40 are very unlikely for the
modern Holocene-era Sun.  Setting significantly stricter quantitative
limits than this for the most energetic solar flares than we have
summarized in Fig.~\ref{fig:2a} requires that we observe a sample of
several dozen very large flares on stars of solar type and of
near-solar age. That, in turn, requires the equivalent of at least
several thousand years of stellar time in the combined observational
sample, to be observed in X-ray, EUV, or optical
emissions. Additional, but less direct, limits could be inferred from
estimated starspot coverages from many thousands of Sun-like stars in,
e.g., observations being made by the Kepler satellite.


\begin{acknowledgments}
We thank the reviewers for helpful questions that improved the
consistency and presentation of our results.
We thank the International Space Studies Institute in Bern, CH, for
support of the team meetings. CJS was supported through the Lockheed
Martin Independent Research and Development program. 
\end{acknowledgments}

\end{article}




%

%
%
%
%
%



\end{document}